\documentclass[twocolumn,notitlepage,nofootinbib,showpacs,preprintnumbers,superscriptaddress,amsmath,amssymb,reprint]{revtex4-1}
\usepackage{graphicx}
\usepackage{dcolumn}
\usepackage{bm}
\usepackage{amsmath}
\usepackage{multirow}
\newcommand{\astra}{{\sc Astra }}

\newcommand{\mean}[1]{\mbox{$\langle{#1}\rangle$}}

\usepackage[colorlinks=true]{hyperref}

\begin{document}

\title{Analysis and Measurement of the Transfer Matrix of a \\ 9-cell, 1.3-GHz Superconducting Cavity} 
\author{A. Halavanau$^{1,2}$, N. Eddy$^2$, D. Edstrom Jr.$^{2}$, E. Harms$^2$, A. Lunin$^2$, P. Piot$^{1,2}$, A. Romanov$^{2}$, J. Ruan$^{2}$, N. Solyak$^2$, V. Shiltsev$^2$ \\
$^1$ Department of Physics and Northern Illinois Center for Accelerator \& \\
Detector Development, Northern Illinois University, DeKalb, IL 60115, USA \\
$^2$ Fermi National Accelerator Laboratory, Batavia, IL 60510, USA} 
\date{\today}
\begin{abstract}
Superconducting linacs are capable of producing intense, stable, high-quality electron beams that 
have found widespread applications in science and industry.  The 9-cell 1.3-GHz superconducting standing-wave accelerating RF cavity originally developed for $e^+/e^-$ linear-collider applications [B. Aunes, {\em et al.} Phys. Rev. ST Accel. Beams {\bf 3}, 092001 (2000)] has been broadly employed in various superconducting-linac designs. In this paper we discuss the transfer matrix of such a cavity and present its measurement performed at the Fermilab Accelerator Science and Technology (FAST) facility. The experimental results are found to be in agreement with analytical calculations and numerical simulations.
\end{abstract}
\preprint{FERMILAB-PUB-17-020-APC}
\pacs{41.75Ht, 41.85.-p, 29.17.+w, 29.27.Bd}
\maketitle
\section{Introduction}
The 1.3-GHz superconducting radiofrequency (SRF) accelerating cavities were originally developed in the context of
 the TESLA linear-collider project~\cite{Aune:2000gb} and were included in the baseline design of the international linear collider (ILC)~\cite{ilc}
and in the design of various other operating or planned accelerator facilities. Projects based on such a cavity include electron-~\cite{LCLS2,ilc}, muon-~\cite{Popovic:2005mj}, 
and proton-beam accelerators~\cite{PIP2} supporting fundamental science 
and compact high-power industrial electron accelerators~\cite{Kephart:SRF2015-FRBA03}. 
Such a cavity is a 9-cell standing-wave accelerating structure operating in the $TM_{010,\pi}$ mode.  
The transverse beam dynamics associated to such a cavity has been extensively explored over  
the last decade and focused essentially on numerical simulations of single-bunch emittance dilution due to the field asymmetries~\cite{Piot:2005id,dohlusEPAC08,luninIPAC2010,vivoli} and multibunch effects due to trapped modes~\cite{hom}. Most recently, experiments aimed at 
characterizing the transverse beam dynamics in this type of SRF cavity were performed~\cite{FNPLNote,Halavanau:2016wax,Halavanau:2016ivl}. 
In this paper we discuss the measurement and  analysis of the transverse transfer matrix of a 9-cell 1.3-GHz SRF cavity. 
In particular, we compare the results with the Chambers' analytical model~\cite{Chambers}.\\

In brief, an analytical model of the transverse focusing in the accelerating cavity can be derived by considering the transverse motion 
of the particle in a standing wave RF field with axial field $E_z(z,t)=E_0\sum_{n}a_n\cos{(nkz)}\sin(\omega t + \phi)$, where $E_0$ is the peak field,
$nk$ is the wave number associated to $n$-th harmonic of amplitude $a_n$, $\phi$ is an arbitrary phase shift, and $z$
is the longitudinal coordinate along the cavity axis. 

The ponderomotive-focusing force is obtained under the paraxial approximation as
 $F_r=-e(E_r-v B_\phi)\approx e r \frac{\partial E_z}{\partial z}$ where $v\simeq c$ is the particle velocity along the axial direction. 
Ref.~\cite{Serafini} shows that the force averaged over one RF-period in the first order of perturbation theory yields the focusing strength,
$\bar{K}_r = - \frac{(E_0 e )^2}{8(\beta \gamma m c^2)^2}$, for the case of a ``pure'' standing wave resonator (where the spatial profile of the axial field is modeled as $E_z(z) \propto \cos(kz)$ inside the cavity) originally considered in Ref.~\cite{Chambers}.
The equation of motion then takes form:
\begin{equation}
\label{EOM}
 x''+\left(\frac{\gamma '}{\gamma}\right)x'+ \bar {K}_r \left(\frac{\gamma '}{\gamma}\right)^2 x=0,
\end{equation}
where $x$ is the transverse coordinate, $x'\equiv \frac{dx}{dz}$, $\gamma' \equiv \frac{d\gamma}{dz} = e E_0 \cos{(\phi)}/m_0 c^2 \equiv \bar{G}_{RF}/m_0 c^2$ 
is the normalized energy gradient, where $\gamma$ is the Lorentz factor.

The solution of the Eq.~\ref{EOM} through the cavity is of the form $\mathbf{x}_f = R \mathbf{x}_i$,
 where $\mathbf{x} \equiv (x,x')^T$, here $R$ is a $2\times2$ matrix,  
and the subscripts $i$ and $f$ indicate upstream and downstream particle coordinates respectively.
According to Chambers' model, the elements of $R$ are given by~\cite{Chambers,Serafini, Rosenzweig, Hartman,Reiche:1997ek}:
\begin{eqnarray} 
\label{Chambers}
\nonumber
R_{11}&=&\cos{\alpha} - \sqrt{2}\cos{(\phi)}\sin{\alpha},\\ \nonumber
R_{12}&=&\sqrt{8}\frac{\gamma_i}{\gamma '}\cos{(\phi)}\sin{\alpha},\\ 
R_{21}&=&-\frac{\gamma '}{\gamma_f}\left[\frac{\cos{(\phi)}}{\sqrt{2}}+\frac{1}{\sqrt{8}\cos{(\phi)}}\right]\sin{\alpha},\\ \nonumber
R_{22}&=& \frac{\gamma_i}{\gamma_f}[\cos{\alpha}+\sqrt{2}\cos{(\phi)}\sin{\alpha}], \nonumber
\end{eqnarray}
where $ \alpha\equiv\frac{1}{\sqrt{8}\cos{(\phi)}}\ln{\frac{\gamma_f}{\gamma_i}}$, 
$\gamma_f\equiv\gamma_i + \gamma' L \cos \phi $ is the final Lorentz factor (where $L$ is the cavity length). 
The determinant associated to the $2\times2$ block of the matrix is $|R|_{2\times2}=\gamma_i/\gamma_f$. 
The latter equation also holds for the vertical degree of freedom $(y,y')$ owing to the assumed cylindrical symmetry.
Under such an assumption the equations for the vertical degree of freedom are obtained 
via the substitutions $x\leftrightarrow y$, $1\leftrightarrow 3$ and $2\leftrightarrow 4$. The total 
transverse transfer matrix determinant is then $|R|_{4\times4}=(\gamma_i/\gamma_f)^2$.

The assumed axially-symmetric electromagnetic field invoked while deriving Eq.~\ref{Chambers} is often violated, e.g., due to asymmetries introduced by the input-power (or forward-power) and high-order-mode (HOM) couplers. The input-power coupler couples the RF power to the cavity while the HOM couplers damp the harmful trapped fields potentially excited as 
long trains of bunches are accelerated in the SRF cavities. In addition to the introduced field asymmetry, the coupler can also impact the beam via geometrical wakefields~\cite{luninIPAC2015,baneEPAC2008}. \\

The measurement of the transverse matrix of a standing wave accelerating structure (a plane-wave transformer, or PWT)
 was reported in Ref.~\cite{Reiche:1997ek} and benchmarked against an ``augmented" Chambers' model detailed in~\cite{Serafini}. This refined model accounts for the presence of higher-harmonic spatial content in the axial field profile $E_z(r=0,z)$. The present paper extends such a measurement to the case of a 1.3-GHz SRF accelerating
 cavity and also investigates, via numerical simulation, the impact of the auxiliary couplers on the transfer matrix of the cavity. These simulations and measurements generally indicate that higher spatial harmonics do not play a significant role for the case of the TESLA cavity. Additionally, we note that the presented measurements are performed in  a regime where the energy gain through the cavity is comparable to the beam injection energy [$\gamma_i \sim \gamma' L$].
In such a regime, the impact of field asymmetries are expected to be important.
%
\section{Numerical Analysis~\label{sec:simu}}
To investigate the potential impact of the couplers, a 3D electromagnetic model of the cavity, including auxiliary couplers, 
was implemented in {\sc hfss}~\cite{hfss}. The simulated 3D electromagnetic field map was imported as an external
 field in the {\sc astra} particle-tracking program~\cite{ASTRAmanual}. The program {\sc astra} tracked particles in the 
presence of external field from first principle via a time-integration of the Lorentz equation. Additionally, {\sc astra} can include space-charge effects using a quasistatic particle-in-cell approach based on solving Poisson's equation in the bunch's rest frame~\cite{ASTRAmanual}. \\

\begin{figure}[hhhhh!!!!!!]
\centering
\includegraphics[width=0.99\linewidth]{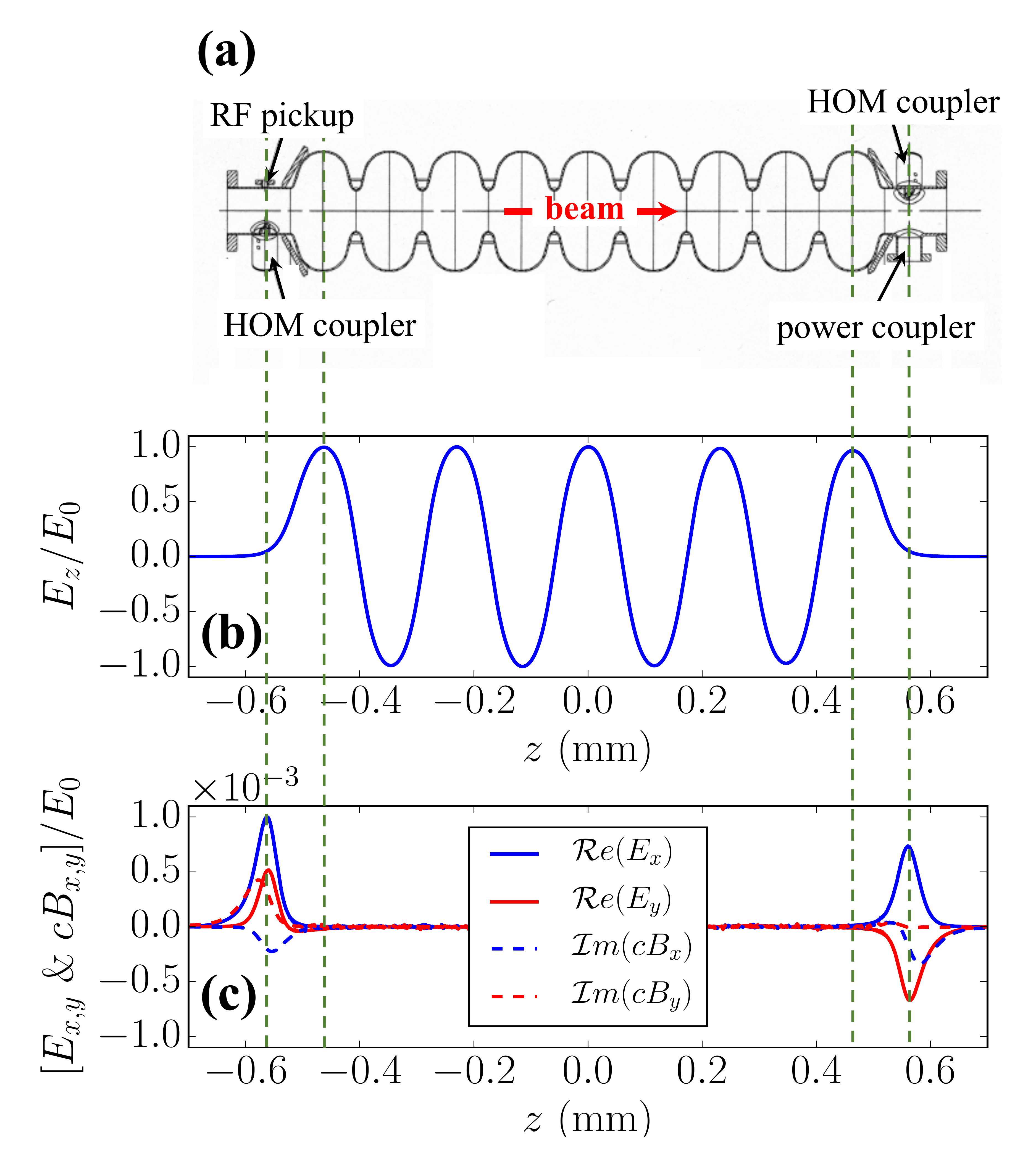}
\caption{\label{fig:cavityhfssmap} Schematics of the TESLA-type cavity considered in the present 
study (a) along with the axial $E_z\equiv E_z(r=0,z)$ (b) and transverse (c) electromagnetic fields simulated on the cavity geometric axis $r=0$.
In (b) and (c) the field are dumped at a time where the electric $E_x, E_y,$ and $E_z$  are real while the magnetic $cB_x$ and $cB_y$ fields are imaginary. All the fields are normalized to the maximum axial electric field $E_0$.}
\end{figure}

The electromagnetic field map $\{\pmb E(x,y,z), \pmb B(x,y,z)\}$  from {\sc hfss} was generated  
over a rectangular computational domain with $x,y\in[- 10,+10]$~mm from the cavity axis and for $z\in[-697.5,+697.5]$~mm with respect to the cavity center along the cavity length; see Fig.~\ref{fig:cavityhfssmap}(a). 
The mesh sizes in the corresponding directions were respectively taken to be  $\delta x =\delta y=0.5$~mm  and $\delta z=1$ mm. The electromagnetic simulations assume a loaded quality factor $Q\simeq 3\times 10^6$ as needed for the nominal ILC operation. Such a loaded $Q$ corresponds to the inner conductor of the input-coupler having a 6-mm penetration depth~\cite{juntong}. Figures~\ref{fig:cavityhfssmap}(b) and (c) respectively present the axial and transverse fields simulated along the cavity axis and normalized to the peak axial field $E_0\equiv \max[E_z(r=0,z)]$. As can be seen in Fig.~\ref{fig:cavityhfssmap}(c) the impact of the coupler, aside from shifting the center of the mode,  also introduces {\em time-dependent} 
transverse electromagnetic fields that will impact the beam dynamics. 
Given the field map loaded in {\sc astra}, the program introduces the time dependence while computing the external Lorentz force experienced by a macroparticle at position $\pmb r\equiv(x,y,z)$ at a given time $t$ as  
\begin{eqnarray}
\pmb F(\pmb r,t) &=& q [ \pmb E(\pmb r)\sin \Psi(t) + \pmb v \times \pmb B(\pmb r)\cos \Psi(t)],
\end{eqnarray}
where $\Psi(t)\equiv \omega t + \phi$ (with $\omega\equiv 2\pi f$ and $f=1.3$ GHz is the frequency) and $q$  and $v$ are respectively the macroparticle charge and velocity. In the latter equation the time origin is arbitrarily selected to ensure $\phi=0$ corresponds to on-crest acceleration. \\

In order to deconvolve the impact of the auxiliary couplers  from the dominant ponderomotive focusing of the cavity, numerical simulations based on a cylindrical-symmetric model were also performed. For these calculations the axial electric field $E_z(r=0, z)$ displayed in Fig.~\ref{fig:cavityhfssmap}(b) is imported in {\sc astra} where the corresponding transverse electromagnetic fields at given positions $(r,\theta,z)$ are computed assuming an ideal TM$_{010}$ mode and under the paraxial approximation as $E_r=-\frac{r}{2}\frac{\partial E_z(r=0, z)}{\partial z}$ and $B_{\phi}=\frac{i \omega r}{2c^2} E_z(r=0, z)$~\cite{helm}. 

\begin{figure}[t]
\centering
\includegraphics[width=1.\linewidth]{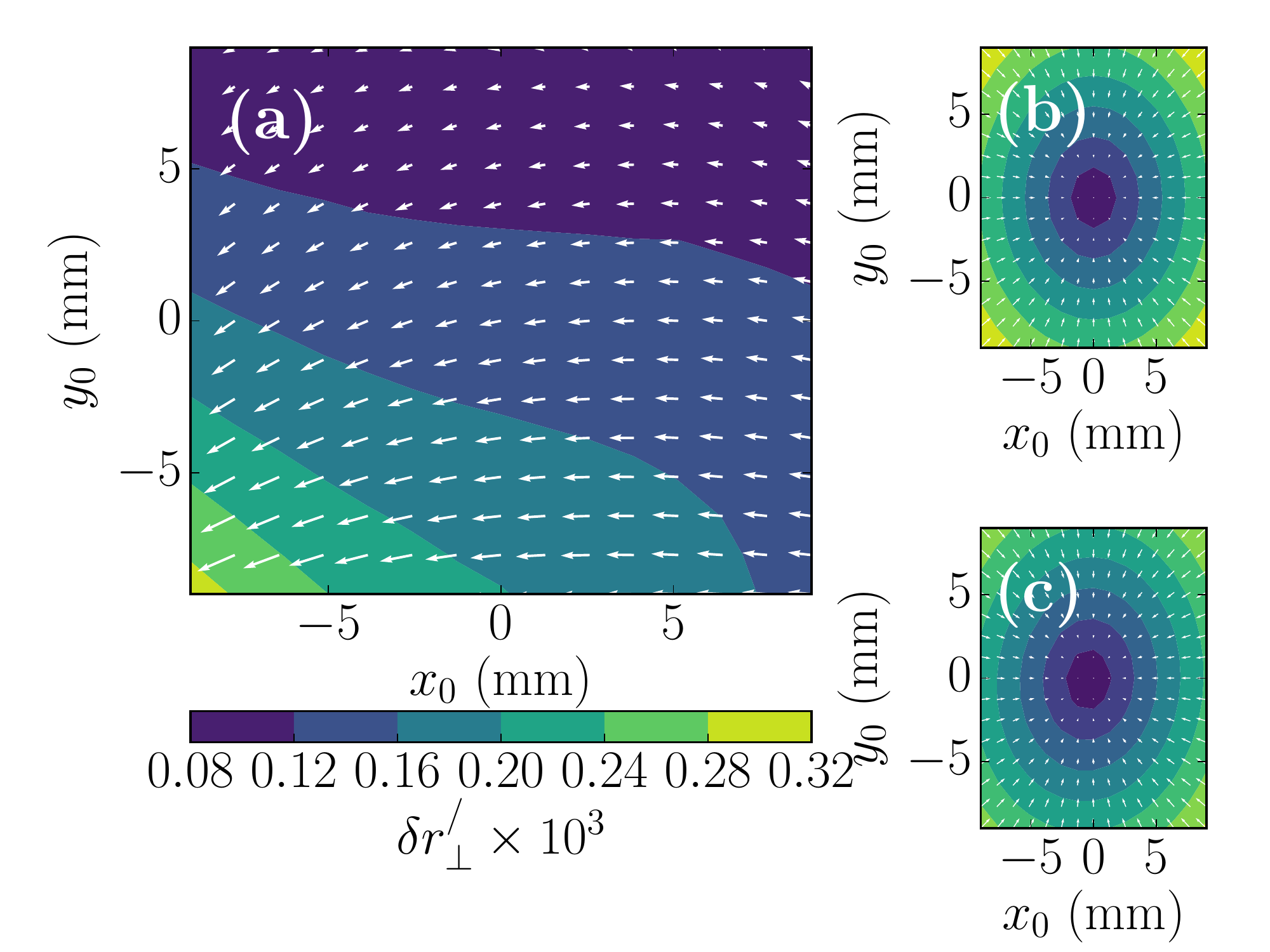}
\caption{\label{fig:momenta} Transverse momentum magnitude (false color contours) and directions (arrows) simulated downstream of the cavity as a function of initial positions. Plot (a) displays the momentum-kick contribution from the auxiliary
 couplers only, i.e.  ${\delta r'}_{\perp}\equiv \frac{1}{\delta P_{\parallel}}|\delta \pmb P_{\perp} - k_p {\pmb  r}_{\perp,0}|$ (see Eq.~\ref{eq:Pexp1} and~\ref{eq:Pexp2}) where $\delta P_{\parallel}$ is the increase in longitudinal momentum, while plots (b) and (c) show respectively the transverse momentum simulated using the cylindrical-symmetric (b) and the 3-D field map (c) models for the cavity. Plot (a) is obtained as the difference between plots (c) and (b). These simulations were performed for 10-MeV electrons with $E_0=30$~MV/m (corresponding to 
$\bar{G}\simeq 15$~MeV/m) and $\phi=0^{\circ}$. }
\end{figure}
\begin{figure}[t]
\centering
\includegraphics[width=1.\linewidth]{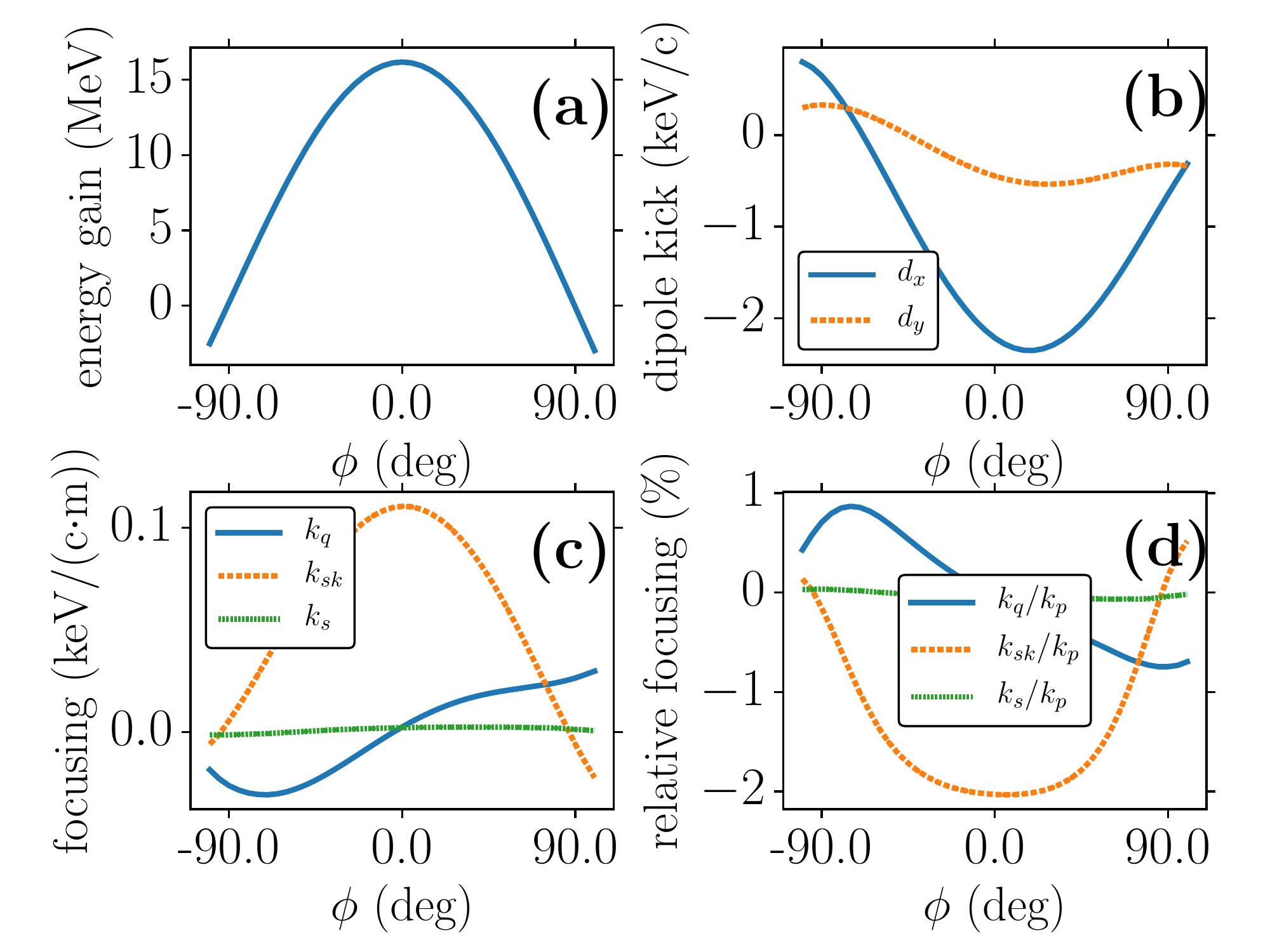}
\caption{\label{fig:kickstrength} Energy gain (a), dipole kicks (b), absolute (c) and relative (d) focusing strengths as function of phase $(\phi=0$ corresponds to on-crest acceleration). The 
relative focusing strength is normalized to the ponderomotive focusing strength $k_p$ in Eq.~\ref{eq:Pexp2}. The simulation conditions are an injection energy of 10 MeV and $E_0=30$~MV/m. }
\end{figure}

In order to quantitatively investigate the transverse beam dynamics in the cavity, we consider a monoenergetic distribution
 of macroparticles arranged on  the vertices of a $2\times2$ transverse grid in the $(x,y)$ plane with
 distribution $\sum_i \sum_j \delta (x-i\Delta x) \delta (y-j\Delta y)$ where $\delta(x)$ is 
Dirac's function and taking $\Delta x =\Delta y =0.3$~mm. The macroparticles, with vanishing incoming transverse momenta and located within the same axial position, are tracked through the cavity field and their final transverse momenta recorded downstream of the cavity. Figure~\ref{fig:momenta}(a) displays the change in transverse momentum $\delta P_{\perp}$ imparted by the auxiliary
 couplers normalized to the change in longitudinal momentum $\delta P_{\parallel}$. This is computed as the difference  between {\sc astra} simulations using the cylindrical-symmetric field [Fig.~\ref{fig:momenta}(b)] from the ones based on the 3D field map [Fig.~\ref{fig:momenta}(c)]. Figure~\ref{fig:momenta}(a) indicates a strong dipole-like field and also hints to the presence of higher-moment components. To further quantify the impact of the auxiliary couplers, we write the change in transverse momentum as an electron passes through the cavity $\delta \pmb P_{\perp} \equiv (\delta p_x,\delta p_y)^T$ as an affine function of the input transverse coordinates $\pmb r_{\perp,0} \equiv (x_0,y_0)^T$ (here the superscript $^T$ represents the transpose operator)
\begin{eqnarray}~\label{eq:Pexp1}
\delta \pmb P_{\perp}  =\pmb d + M \pmb  r_{\perp,0},
\end{eqnarray}
where  $\pmb d\equiv(d_x, d_y)$ is a constant vector accounting for the dipole kick along each axis, and $M$ is a $2\times2$ correlation matrix. The latter equation can be rewritten to decompose the final momentum in terms of the strength characterizing the various focusing components~\cite{Li:1993}
\begin{eqnarray}~\label{eq:Pexp2}
\begin{pmatrix} \delta p_x  \\ \delta p_y \end{pmatrix} &=& \begin{pmatrix} d_x \\ d_y \end{pmatrix} + k_p \begin{pmatrix} x_0 \\ y_0 \end{pmatrix}  + k_q \begin{pmatrix} x_0  \\ -y_0  \end{pmatrix}   \nonumber \\
 && + k_{sk} \begin{pmatrix} y_0  \\ x_0  \end{pmatrix} + k_{s} \begin{pmatrix} y_0  \\ -x_0  \end{pmatrix},
\end{eqnarray}
where  $k_{p,q} \equiv (M_{11}\pm M_{22})/{2}$, and  $k_{sk,s} \equiv (M_{12}\pm M_{21})/{2}$  respectively account for the axially-symmetric ponderomotive, quadrupole, skew-quadrupole and solenoidal focusing effects. It should be pointed out that the coefficients introduced in the latter equation are implicit functions of the cavity field and operating phase.
Furthermore, the linear approximation resulting in Eq.~\ref{eq:Pexp1} requires validation.
In order to find the focusing strength we performed simulations similar to the one presented in Fig.~\ref{fig:momenta}(c) and directly compute the offset $\pmb d$ and correlation matrix $M$ necessary to devise the focusing strengths in Eq.~\ref{eq:Pexp2}. Such an analysis was implemented to provide the steering and focusing strength as a function of the injection phase $\phi$ as summarized in Fig.~\ref{fig:kickstrength}.  Our analysis confirms the presence of higher-moment components such as quadrupole and skew-quadrupole terms as investigated in Ref.~\cite{dowel}. It also indicates the strength of these 
quadrupolar components is very small compared to the cylindrical-symmetric ponderomotive focusing,
 specifically $k_{sk}\sim k_q \sim {\cal O} (10^{-2} \times k_p)$. Finally, we observe that the 
solenoidal contribution $k_s \sim  {\cal O} (10^{-4} \times k_p)$ is insignificant.  The relatively weak focusing strength arising from the presence of the auxiliary couplers confirm that the transfer matrix will be essentially dominated by the ponderomotive focusing. Therefore we expect the couplers to have negligible impact on the transfer-matrix measurement reported in the next Section. It should however be noted that the time dependence of these effects, especially of the dipole kick, can lead to significant emittance increase via a head-tail effect where different temporal slice within the bunch experience a time-varying kick resulting in a dilution of transverse emittance. Such an effect is especially important when low-emittance low-energy beams are being accelerated in a string of cavities~\cite{Piot:2005id,luninIPAC2010,vivoli}. 

\section{Experimental setup  \& Method}
\begin{figure}[t]
\centering
\includegraphics[width=0.96\linewidth]{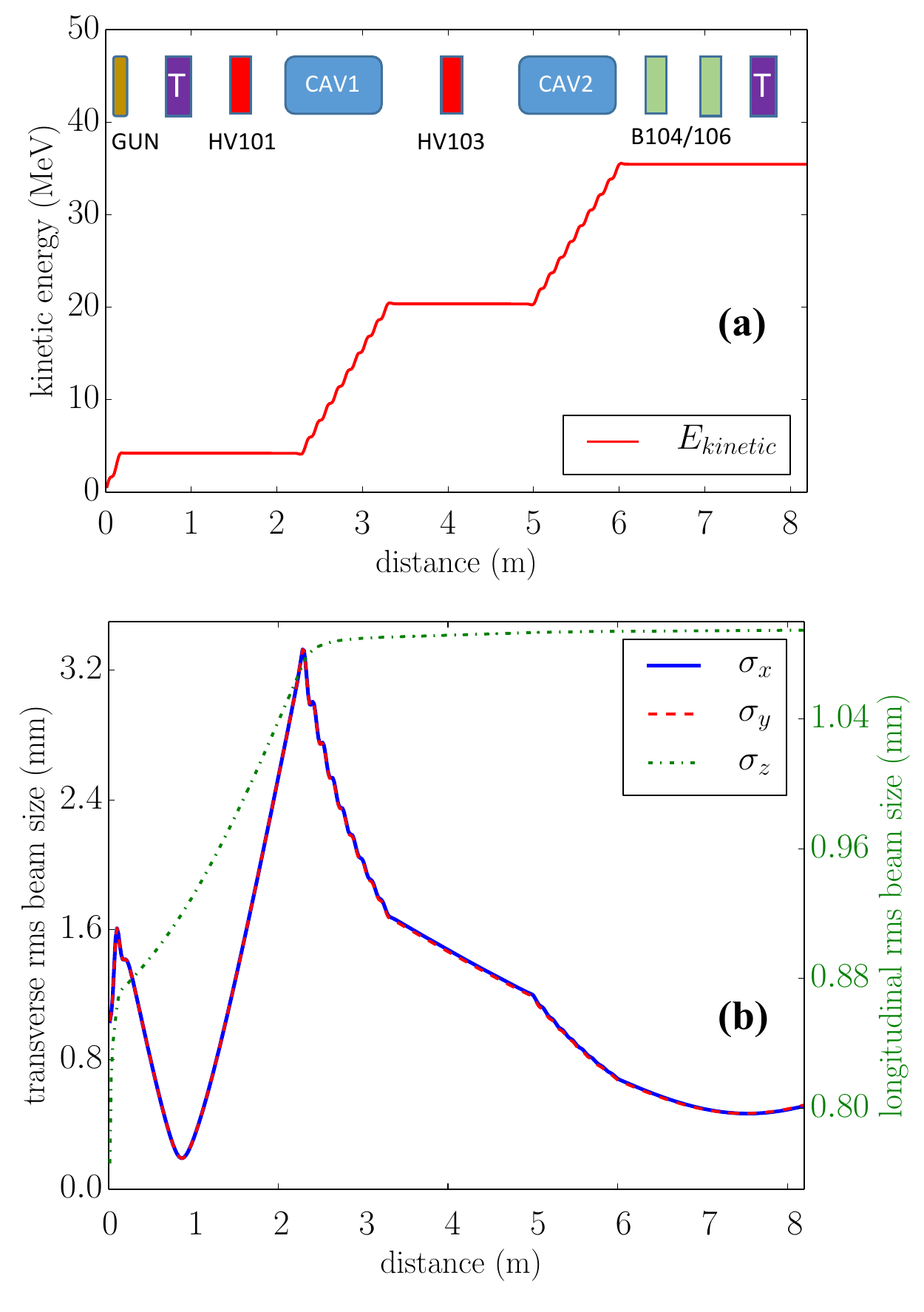}
\caption{\label{fig:exp} Experimental setup under consideration and associated energy gain
 (a) and transverse and longitudinal bunch sizes (b) simulated with \astra. In the diagram displayed in (a),  the labels ``CAV" , ``T", ``HV'', and ``B" respectively correspond to the SRF cavities, the integrated-current monitors (ICM), the magnetic steerers, and beam-position monitors.}
\end{figure}

The experiment was performed in the electron injector of the IOTA/FAST facility~\cite{FAST, FAST2}. The experimental setup is diagrammed in Fig.~\ref{fig:exp}(a). In brief, an electron beam photoemitted from a high-quantum-efficiency semiconductor 
photocathode is rapidly accelerated to $\sim 5$~MeV in a L-band $1+\frac{1}{2}$ cavity radiofrequency (RF) gun. 
The beam energy is subsequently boosted using two 1.03 m long 1.3-GHz SRF accelerating cavities [labeled as CAV1 and CAV2 in Fig.~\ref{fig:exp}(a)] 
up to maximum of $\sim 52$~MeV. In the present experiment the average accelerating gradient of the accelerating cavities was respectively set to 
 $\bar{G}_{CAV1}\simeq 15$~MeV/m and $\bar{G}_{CAV2}\simeq 14$~MeV/m. The simulated bunch transverse sizes and length along the IOTA/FAST 
photoinjector appear in Fig.~\ref{fig:exp} for the nominal bunch charge ($Q=250$~pC) and settings used in the experiment. The corresponding peak current, $\hat{I}\simeq 30$~A, is small enough to ensure wakefield effects are insignificant $-$ from From Fig.~4 of Ref.~\cite{baneEPAC2008} we estimate the transverse geometric wakefield to yield a kick on the order of 1~eV/$c$, i.e., two order of magnitude lower than the dipole kick given in Fig.~\ref{fig:kickstrength} over the range of phase $\phi\in[-30^{\circ},30^{\circ}]$. The simulated kinetic energy downstream of CAV2 is $K\simeq 34$~MeV consistent with the measured value. \\

The available electron-beam diagnostics include cerium-doped yttrium aluminum garnet (Ce:YAG) scintillating crystals for transverse beam size measurement upstream of CAV1 and downstream of CAV2 and beam position monitors (BPMs) which were the main diagnostics used duing our experiment. Each BPM consists of four electromagnetic pickup ``button" antennae located $90^{\circ}$ apart at the same axial position and at a radial position $35$-mm from the beamline axis. The beam position $u = (x,y)$ is inferred from the beam-induced voltage on the antenna using a 7-th order polynomial $u=\sum_i a_{u,i} F(\Phi_j)$ where $\Phi_j$ ($j=1,2,3,4$) are the induced voltages on each of the four BPM antenna and the coefficients $a_{u,i}$ are inferred from a lab-bench calibration procedure using a wire-measurement technique; see Ref. \cite{McCrory:2013eta}. At the time of our measurements, the BPM system was
 still being commissioned and the resolution was about $ \simeq 80$~$\mu$m in both dimensions~\cite{Plopez}.

As the starting point of the transfer-matrix measurement, the beam was centered through both cavities CAV1
 and CAV2 using a beam-based alignment procedure. The beam positions $(x_i,y_i)$ [where $i=1,2$] downstream of the CAV2 were recorded for two phase settings ($\phi_{1,2}=\pm30^{\circ}$) and the function $\chi = \sqrt{(x_1-x_2)^2+(y_1-y_2)^2}$ quantifying the relative beam displacement was evaluated.  The settings of the dipole correctors upstream of the cavity CAV2 were then employed as free variables to minimize $\chi$ using a conjugate-gradient algorithm.\\

%

In order to measure the transfer matrix, we used a standard difference-orbit-measurement technique where beam-trajectory perturbations are applied with magnetic steerers located upstream of CAV2 and resulting changes are recorded downstream of the cavity with a pair of BPMs. In our experiment, the perturbations were applied using two sets of horizontal and vertical magnetic steerers (HV101 and HV103) with locations displayed in Fig.~\ref{fig:exp}(a). Orbit perturbations were randomly
generated to populate a large range of initial conditions in the 4D trace space $\mathbf{X_i}\equiv (x_i,x_i', y_i, y_i')$. Only the perturbations for which the beam was fully-transmitted were retained [the charge transmission is inferred from two integrated-current
 monitors (ICM) shown in Fig.~\ref{fig:exp}(a)].
For each measured cavity phase point, 20 different sets of perturbations (associated to a set of upstream dipole-magnet settings) were impressed. The beam was then propagated through CAV2 up to a pair of downstream electromagnetic button-style BPMs. 
The measurement of beam position with CAV2 ``off'' and ``on'', where ``off'' means zero accelerating gradient, 
(indirectly) provided the initial $\mathbf{ X_i}$ and final $\mathbf{ X_f}$ beam positions and divergences respectively upstream and downstream of CAV2. 

Correspondingly, given the $4\times 4$ transfer matrix of the cavity  $R$,  these vectors are 
related via $\mathbf{ X_f}=  R\mathbf{ X_i}$. An initial perturbation $\delta \mathbf{ X_{0_i}}$ to the nominal
 orbit $\mathbf{ X_{0_i}}$ such that $\mathbf{X_i}=\mathbf{ X_{0_i}}+\delta \mathbf{ X_{0_i}}$ will result in an orbit change downstream of CAV2 given by 
\begin{eqnarray}\label{eq:difford}
 \delta \mathbf{ X_{0_f}} = R \delta \mathbf{ X_{0_i}}. 
\end{eqnarray}
Therefore any selected orbit can serve as a reference orbit to find the transformation $R$, assuming the set of perturbed trajectories around this reference is transformed linearly
 (which is the essence of the paraxial approximation). Consequently, impressing a set of $N$ initial perturbations $\delta \mathbf{X}^{(n)}_{0_i}$ where  $n=[1...N]$ results in a system of $N$ equations similar to Eq.~\ref{eq:difford} which can be casted in the matrix form 
\begin{eqnarray}
\Xi_f=R \Xi_i,
\end{eqnarray}
where $\Xi_j$ ($j=i, f$) are $4 \times N$ matrices containing the positions and divergence associated to the $N$ orbit perturbations. This system can then be inverted via a least-squares technique to recover $R$.  \\

The error analysis includes statistical fluctuations (which arise from various sources of jitter) 
and uncertainties on the beam-position measurements. The statistical error bars were evaluated using an analogue of a 
boot-strapping technique. Given that the transformation (\ref{eq:difford}) is linear, 
any couple of initial $\mathbf{X_{k,i}}$ and final $\mathbf{X_{k,f}}$ beam position 
measurements can define the reference orbit while the other couples $(\mathbf{X_{j,i}}, \mathbf{X_{j,f}})$ 
for $j\in N\neq k$ are taken as perturbed orbits and the transfer matrix can be inferred. Consequently, we retrieved 
the transfer matrix $R_j$ associated to a reference orbit $(\mathbf{X_{k,i}}, \mathbf{X_{k,f}})$.  Such a procedure is 
repeated for all orbits $k\in[1,N]$ and the resulting transfer matrix  $R_k$ is recorded. A final 
step consists in computing the average $\mean{R}$ and variance $\sigma_R^2=\mean{R^2-\mean{R}^2}$ over the $N$
 realizations of $R_j$. Finally, the measured value is reported as $R= \mean{R} \pm 2\sigma_R$. 

\section{Experimental results}

The elements of the transfer matrix were measured for nine values of phases in the range $\phi \in  [-20^{\circ},20^{\circ}]$ 
around the maximum-acceleration (or ``crest") phase corresponding to $\phi=0^{\circ}$.
 
For each set of perturbation the beam positions along the beamline were recorded over 4 shots to account for
 possible shot-to-shot variations arising from beam jitter or instrumental error. The corresponding set of 80 orbits were subsequently 
used in the analysis algorithm described in the previous Section. 

\begin{figure}[t]
\centering
\includegraphics[width=1.0\linewidth]{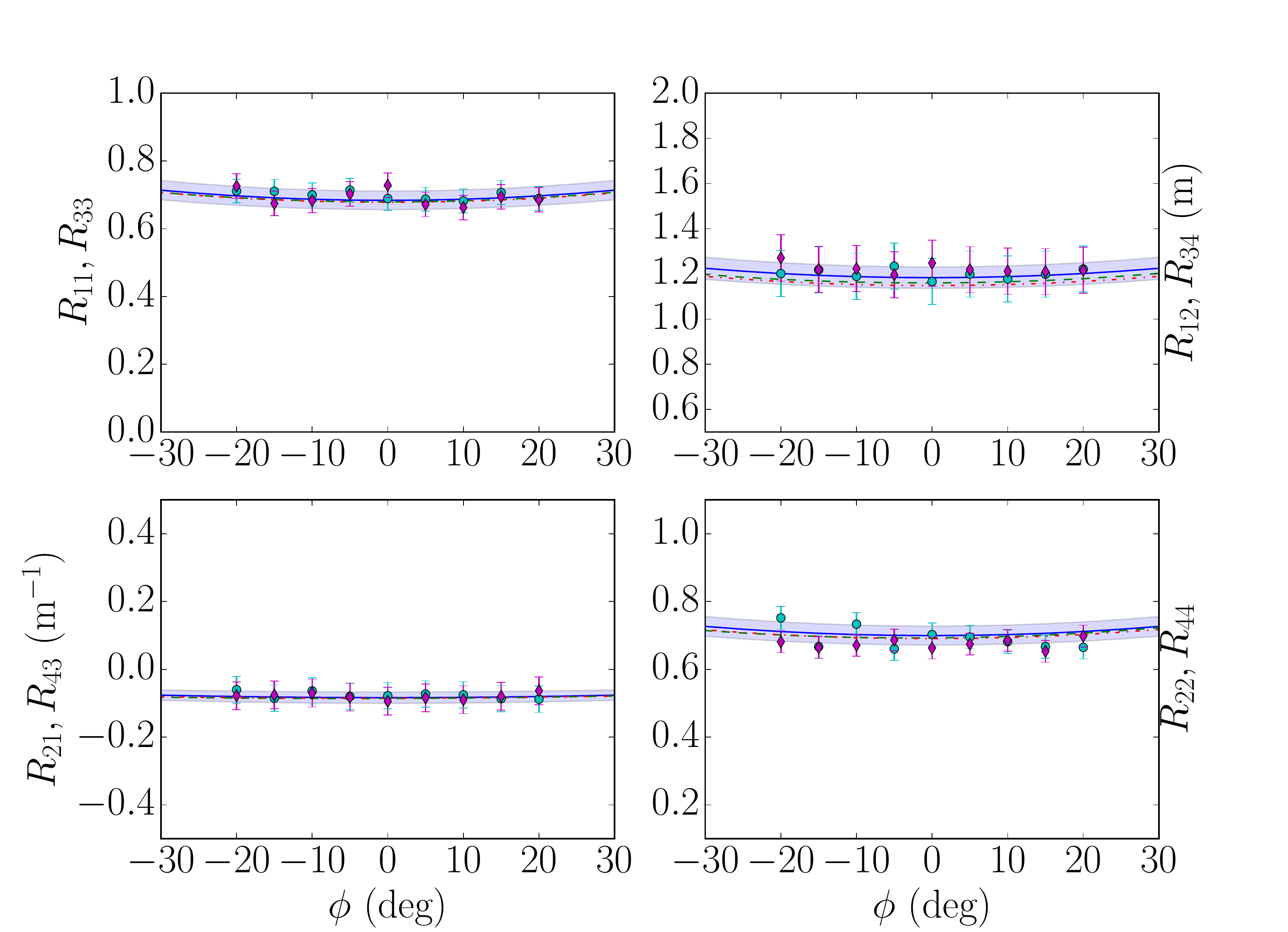}
\includegraphics[width=1.0\linewidth]{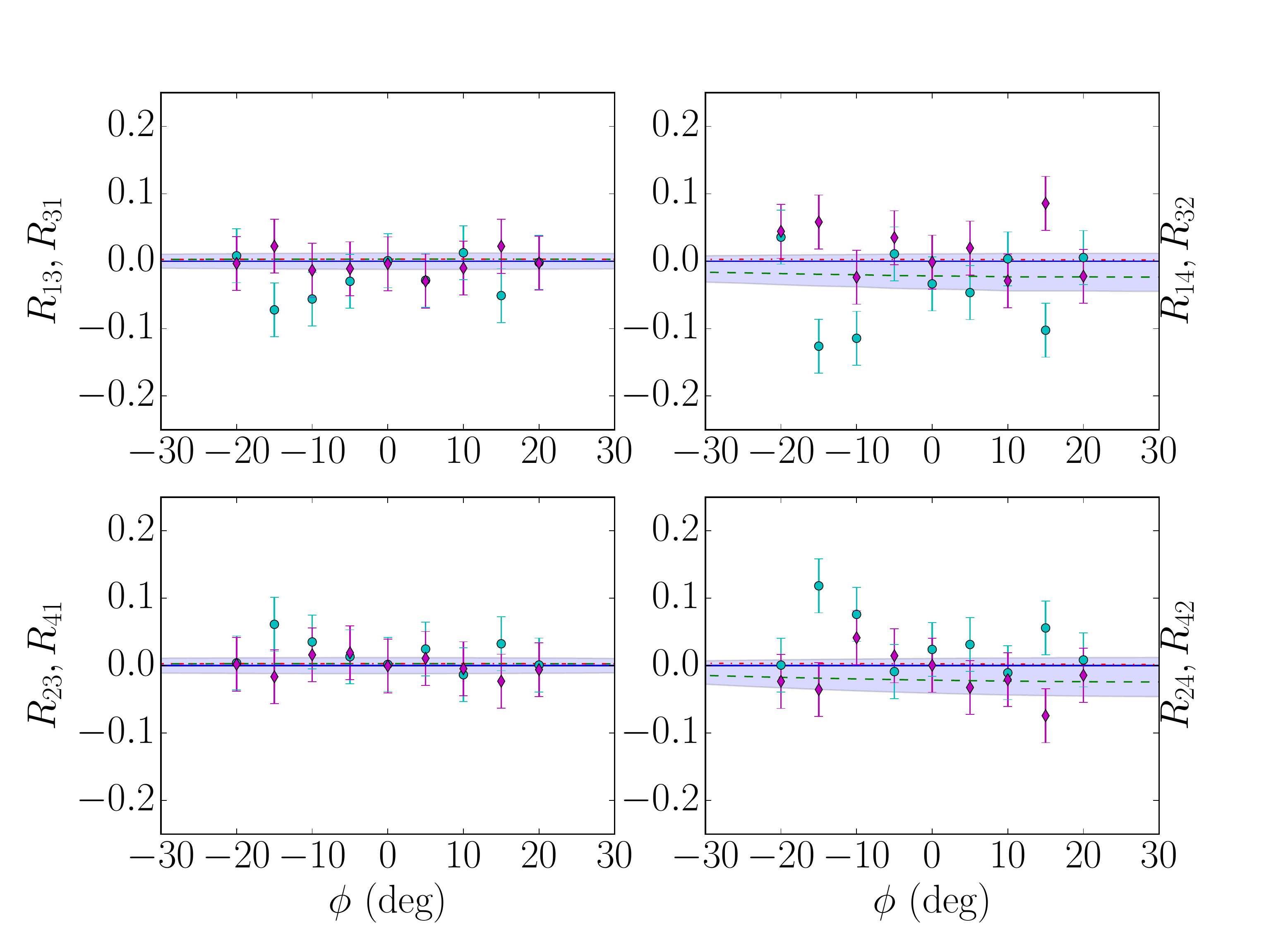}
\caption{\label{mainblock} Diagonal (top four plots) and anti-diagonal (bottom four plots) blocks of the transport matrix. 
The solid (blue) lines represent Chambers' approximation, dashed (green/red) lines are obtained from 3D field map simulations
for $(x,x')$ and $(y,y')$ planes respectively, circular
 markers and purple lozenges correspond to experimental values for $(x,x')$ and $(y,y')$ planes respectively. Shaded area
represents matrix element variation due to RF calibration uncertainties (simulation).}
\end{figure}

The comparison of the recovered transfer matrix elements with the Chambers' model along with the matrix inferred from particle 
tracking with {\sc astra} appear in Fig.~\ref{mainblock}. The shaded areas in Fig.~\ref{mainblock} and subsequent figures
 correspond to the simulated uncertainties given the CAV2 cavity gradient $\bar{G}_{CAV2}=14 \pm 1$~MeV/m. \\

\begin{figure}[t!!!!]
\centering
\includegraphics[width=0.96\linewidth]{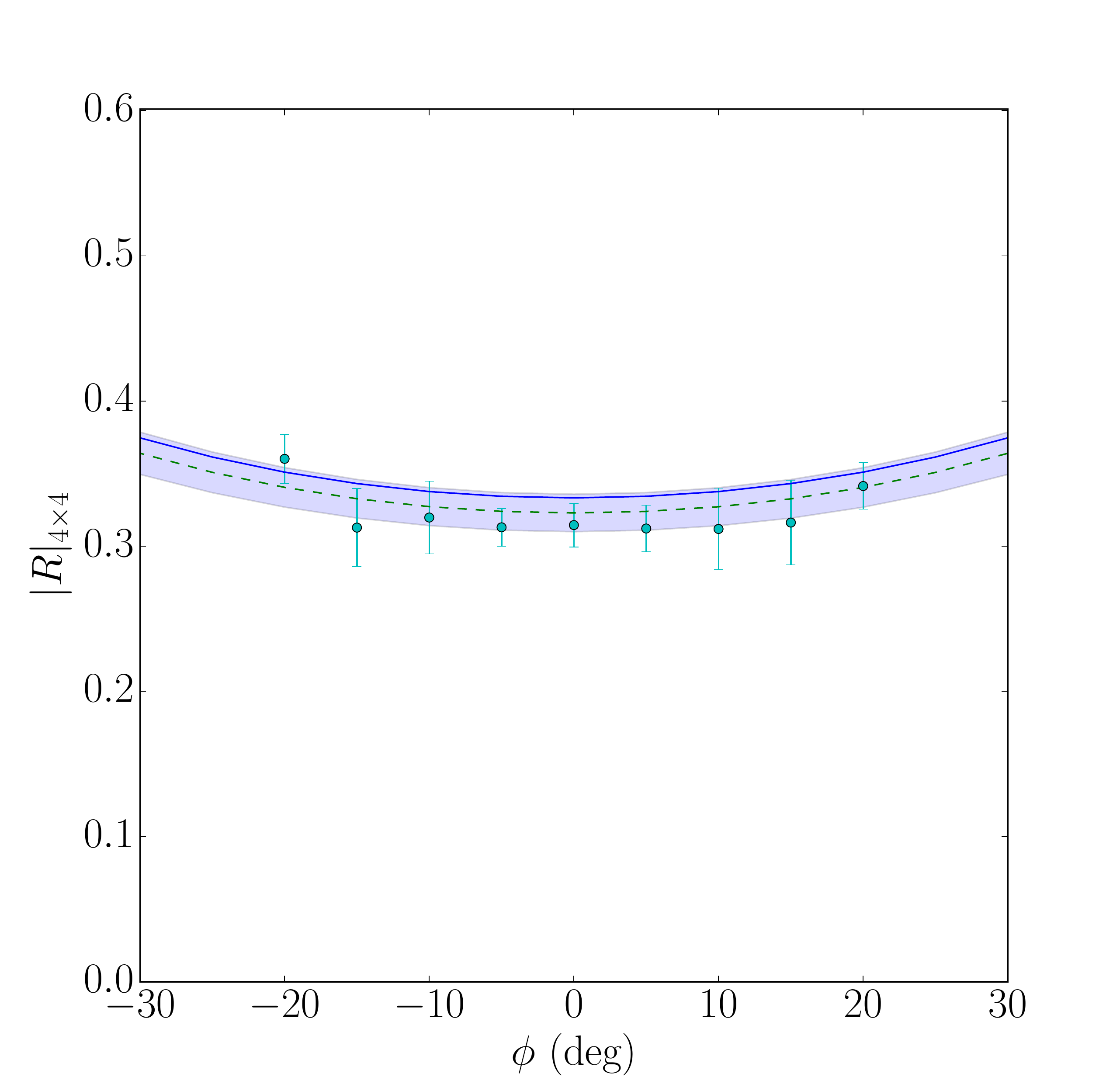}
\caption{ \label{determinants} Measured $4\times4$ transfer-matrix determinant (symbols) compared with the 
Chambers' approximation (solid line) and numerical simulations using the 3D field map (dashed line).  The shaded area represents the uncertainty on the simulations due to RF calibration uncertainties.}
\end{figure}

Overall, we note the very good agreement between the measurements, simulations, and theory. The slight discrepancies between the Chambers' model and the experimental results do not appear to have any correlations and are attributed to the instrumental jitter of the BPMs, RF power fluctuations, cavity alignment uncertainties, halo induced by non-ideal laser conditions. 
During the measurement, we were unable to set the phase of the CAV2 beyond the aforementioned range 
as it would require a significant reconfiguration of the IOTA/FAST beamline. Nevertheless we note that this range of phases is of interest to most of the project currently envisioned.\\

The elements of coupling (anti-diagonal) $2\times2$ blocks of the $4\times4$ matrix, modeled in the simulation are about one order of magnitude smaller than the elements of the diagonal block.
For instance, considering the $x$ coordinate we
find that $R_{13}/R_{11} \sim {\cal O}(10^{-2})$ and $R_{14}/R_{12}\sim {\cal O}(10^{-2})$. This finding corroborates with our experimental results which indicate that $R_{13}/R_{11} \lesssim 0.1$ and $R_{14}/R_{12} \lesssim 0.1$; see Fig.~\ref{mainblock}. The latter observation confirms that, for the range of parameters being explored, the 3D effects associated to the presence of the couplers has small impact on the single-particle beam dynamics as already discussed in Sec.~\ref{sec:simu}.  The measured matrix elements were used to infer the determinant $|R|$ which is in overall good agreement with the simulation and Chambers' models; see Fig.~\ref{determinants}. \\

\begin{figure}[tttt]
\centering
\includegraphics[width=0.95\linewidth]{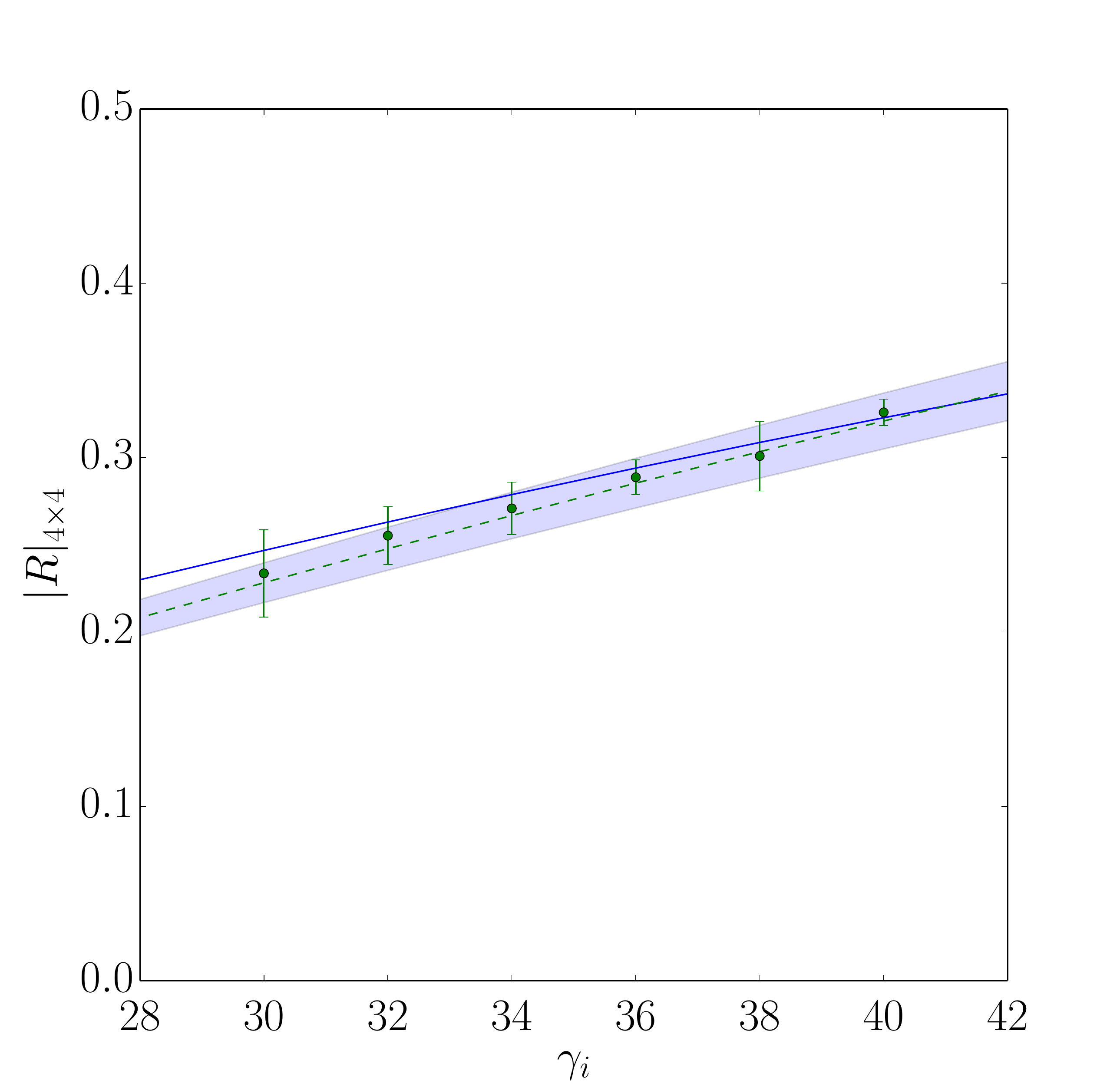}
\caption{ \label{damping} Measured scaling for the $4\times4$ transfer matrix determinant as a 
function of injected-beam Lorentz factor $\gamma_i$ (symbols) compared 
with the Chambers' approximation (solid line) and numerical
simulations using the 3D field map (dashed line).  The shaded area represents the uncertainty on the simulations due to RF calibration uncertainties.}
\end{figure}

Finally, the field amplitude in CAV1 was varied, thereby affecting the injection energy in CAV2 and the transfer matrix element of CAV2 measured. Since the beam remained relativistic the change did not affect the injection phase in CAV2. The resulting determinant (for the $2\times2$ matrix) is expected to follow an adiabatic scaling $\gamma_i/\gamma_f$. The experimental measurement presented in Fig.~\ref{damping} confirm a scaling in $(\gamma_i/\gamma_f)^2$ as expected for the determinant of the $4\times4$ transfer matrix.

\section{Discussion}

In summary, we have measured the transfer matrix of a 1.3-GHz SRF accelerating cavity at IOTA/FAST facility. The measurements are found to be in good agreement with numerical simulations and analytical results based on the Chambers' model. 
In particular, the contributions from the auxiliary couplers are small and does not affect the $4\times 4$ matrix which can be  approximated by a symmetric $2\times 2$-block diagonal matrix within our experimental uncertainties. 
Furthermore the electromagnetic-field deviations from a pure cylindrical-symmetric TM$_{010}$ mode do not significantly affect the {\em single-particle} beam dynamics. 

It should however be stressed that nonlinearities along with the time-dependence of the introduced dipole, 
and non-cylindrical-symmetric first order perturbations contribute to transverse-emittance dilutions~\cite{SainiIPAC10,luninIPAC2010}. Investigating such effects would require beams with ultra-low emittances. A unique capability of the IOTA/FAST photoinjector is its ability to produce flat beams -- i.e. beams with  large transverse-emittance ratios~\cite{piotFB,zhu}. The latter type of beams could produce sub-$\mu$m transverse emittances along one of the transverse dimensions thereby providing an ideal probe to quantify the emittance dilution caused by the cavity's auxiliary couplers.

\section{Acknowledgments}
We are grateful to D. Broemmelsiek, S. Nagaitsev, A. Valishev and the rest of the IOTA/FAST group for their support. This work was partially funded by the US Department of Energy (DOE) under contract DE-SC0011831 with Northern Illinois University.  Fermilab is operated by the Fermi Research Alliance, LLC for the DOE under contract DE-AC02-07CH11359.

\end{document}